\documentclass[11pt,showpacs,preprintnumbers,amsmath,amssymb,prd,nofootinbib,superscriptaddress]{revtex4-2}

\usepackage{dcolumn}
\usepackage{bm}
\usepackage{ifpdf}
\usepackage{hyperref}
\usepackage{bm}
\usepackage{xcolor,color,graphicx,graphics,physics}
\usepackage[spanish,english]{babel}
\usepackage[latin1]{inputenc}
\usepackage[OT1]{fontenc}
\usepackage{latexsym,amssymb,amsmath,amsfonts, slashed}
\usepackage{makeidx}
\usepackage{epsfig,subfigure}
\usepackage{natbib}
\usepackage{epstopdf}
\usepackage{mathrsfs}
\usepackage{hyperref}
\hypersetup{colorlinks=true, linkcolor=blue, citecolor=blue, urlcolor=blue}
\usepackage{enumerate}
\usepackage{cancel}

\usepackage{braket}
\usepackage{fixmath}

\everymath{\displaystyle}
\usepackage{graphicx}

\usepackage[T1]{fontenc}
\usepackage{amsmath}
\usepackage{amssymb}
\usepackage{graphicx}
\usepackage{xcolor}

\newcommand{\bea}{\begin{eqnarray}}
\newcommand{\eea}{\end{eqnarray}}

\newcommand{\til}{\text{\textasciitilde}}
\ExplSyntaxOn
\NewDocumentCommand{\RN}{m}
 {
  \textup{ \int_to_Roman:n { #1 } }
 }
\ExplSyntaxOff

\newcommand{\orcid}[1]{\href{https://orcid.org/#1}{\includegraphics[width=10pt]{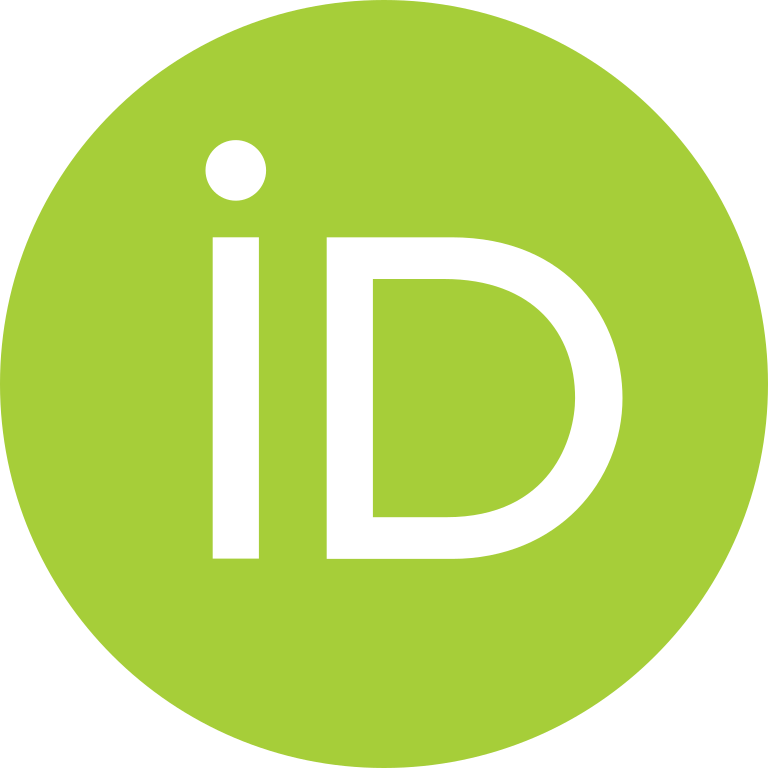}}}

\begin{document}

\title{Gravitational Lorentz-violating $e^-+e^+\to\ell^-+\ell^+$ scattering}

\author{L. A. S. Evangelista \orcid{0009-0002-3136-2234}}
\email{lucassouza@fisica.ufmt.br }
\affiliation{Programa de P\'{o}s-Gradua\c{c}\~{a}o em F\'{\i}sica, Instituto de F\'{\i}sica,\\ 
Universidade Federal de Mato Grosso, Cuiab\'{a}, Brasil}


\author{A. F. Santos \orcid{0000-0002-2505-5273}}
\email{alesandroferreira@fisica.ufmt.br}
\affiliation{Programa de P\'{o}s-Gradua\c{c}\~{a}o em F\'{\i}sica, Instituto de F\'{\i}sica,\\ 
Universidade Federal de Mato Grosso, Cuiab\'{a}, Brasil}



\begin{abstract}

We investigate the gravitational $e^-+e^+\to\ell^-+\ell^+$  scattering process within the framework of gravitoelectromagnetism, a weak-field approximation of gravity analogous to Maxwell's theory of electromagnetism. This process involves the interaction between a fermion and an antifermion mediated by graviton exchange. We consider the nonminimal gravitational sector of the standard model extension and calculate the corrections to the scattering cross section arising from Lorentz violation. The analysis is carried out in two scenarios: (i) at zero temperature and (ii) at finite temperature. To incorporate thermal effects, we employ the thermo field dynamics formalism, which allows for a consistent treatment of quantum fields at finite temperature. The results provide insights into how Lorentz-violating and thermal corrections influence gravitational interactions, particularly relevant in high-energy or astrophysical environments.

\end{abstract}

\maketitle

\section{Introduction}

The Standard Model Extension (SME) expands the Standard Model of particle physics to provide a unified description of all fundamental forces, including gravity. It systematically includes all possible Lorentz- and CPT-violating terms that couple to fields from both the Standard Model and general relativity. The SME is based on the idea that these symmetries, which are exact in the Standard Model, might be broken--either spontaneously or explicitly--at very high energy scales, such as those expected in quantum gravity. As a result, the SME works as an effective field theory that looks for possible low-energy traces of broken symmetries. This idea is also supported by string theory, where spontaneous Lorentz symmetry breaking can happen naturally, giving a theoretical reason to study small deviations from exact symmetry. \cite{colladay1997lorentz, colladay1998lorentz, kostelecky2002signals, kostelecky2004gravity, kostelecky2009electrodynamics, bailey2015short, kostelecky2016testing, kostelecky1989phenomenological, kostelecky1991photon, kostelecky1989spontaneous}.

Building on this foundation, numerous studies have explored the consequences of symmetry breaking within both the minimal and nonminimal sectors of the SME. In the minimal SME, where renormalization remains under control, various works have explored different sectors, including the CPT-even sector \cite{borges2014field, martin2016casimir, frank2006casimir, casana2010parity}, the CPT-odd sector \cite{carroll1990limits, kharlanov2010casimir, martin2017local, altschul2007cerenkov}, the fermion sector \cite{kostelecky1999nonrelativistic, kostelecky2001stability, xiao2016spectra}, and quantum electrodynamics (QED) \cite{schreck2012analysis, santos2018fermion, ferrari2022one}. In these cases, the Lagrangian is modified by introducing symmetry-breaking operators of mass dimension $d \leq 4$. In contrast, the nonminimal SME considers higher-dimensional, nonrenormalizable operators with $d > 4$. Investigations in this sector include studies of the fermion and photon sectors \cite{kostelecky2013fermions, kostelecky2009electrodynamics}, radiative corrections \cite{borges2014generation, borges2016generation, altschul2019bounds}, and Lorentz-violating effects from boundary conditions \cite{borges2020new}, among others. In this work, we investigate Lorentz violation in the gravity sector by introducing a fifth-order coefficient that exhibits characteristics similar to the parity-odd terms found in the QED sector \cite{bailey2015short, kostelecky2016testing}. This background is part of the nonminimal sector of the SME \cite{kostelecky2019gauge, borges2022external}. We consider its coupling in the context of the gravitational scattering process $ e^- + e^+ \to \ell^- + \ell^+ $, which involves interactions between gravitons and fermions.

In the standard QED, the $e^-+e^+\to \ell^-+\ell^+$ scattering process describes the interaction between a fermion and an antifermion, mediated by a virtual photon. This type of fermion-antifermion scattering has been extensively studied in the literature \cite{sanna1998differential, kauppila1989comparisons, nahar2020positron, arbuzov1997small, bufalo2014causal}, including works that consider Lorentz-violating effects \cite{casana2012effects, fu2016moller, charneski2012lorentz, cabral2024exploring}. Building on these studies, we extend the analysis to the gravitational counterpart of this process, where the interaction is mediated by gravitons instead of photons. This investigation is carried out within the framework of Gravitoelectromagnetism (GEM).

GEM is a weak-field approximation of gravity that treats gravitational interactions in a manner analogous to Maxwell's theory of electromagnetism \cite{faqir2010lagrangian, bakopoulos2016gravitoelectromagnetism, mashhoon2003gravitoelectromagnetism, costa2008gravitoelectromagnetic, caceres2025gravitoelectromagnetismkerrschildweyldouble}.One common method for formulating GEM involves decomposing the Weyl tensor into its gravitoelectric and gravitomagnetic components {\cite{ramos2006derivation, ramos2006differential, ramos2018weyl, ramos2020abelian}}, {commonly referred to in the literature as Weyl GEM}.
 This decomposition allows for the construction of a Lagrangian suitable for quantization, leading to the interpretation of the graviton as a spin-2 particle \cite{faqir2010lagrangian}. In this formalism, the key quantity is the potential tensor $A_{\mu\nu}$, which encodes all information about the gravitoelectric and gravitomagnetic fields. The structure of GEM closely resembles that of electromagnetism, which is expected since its formulation is inspired by QED, though it involves higher-rank tensors due to the spin-2 nature of the graviton. A comparative analysis of scattering processes in GEM and QED is presented in Ref. \cite{evangelista2025gravitational}. Moreover, several other studies have explored scattering phenomena and related physical processes within the GEM framework \cite{ santos2017gravitational, alesandrogravitacional, Casimir, bailey2010lorentz}.

It is important to note that GEM is only applicable in the weak-field limit, which means it cannot describe strong gravitational systems like black holes or pulsars themselves. However, it can be used to explore residual effects on nearby objects. From the perspective of quantum field theory, the {Weyl GEM} formalism allows not only the study of interactions between gravitons and fermions but also the investigation of how these interactions behave under thermal effects \cite{caceres2025gravitoelectromagnetismkerrschildweyldouble}. To address this, we adopt the Thermo Field Dynamics (TFD) approach.

TFD is a formalism in quantum field theory that incorporates thermal effects into the system \cite{Umezawa1, Umezawa2, arimitsu1987non, Book, Umezawa22, Khanna1, Khanna2, Santana1, Santana2}. These effects emerge from the need to represent statistical averages as expectation values of quantum operators. To achieve this, TFD extends the conventional Hilbert space by introducing a duplicate (tilde) space, resulting in a total space given by $\mathcal{H}_T = \mathcal{H} \otimes \tilde{\mathcal{H}}$, where $\mathcal{H}$ is the original Hilbert space and $\tilde{\mathcal{H}}$ is its tilde counterpart. As a result, field operators must be redefined using Bogoliubov transformations, which express them as linear combinations of operators acting on both spaces. Although other methods exist for incorporating temperature--such as the Matsubara formalism \cite{matsubara} and the Closed-Time Path (CTP) approach \cite{schwinger1961brownian}--TFD is chosen here for its simpler and more direct application to tree-level scattering processes. For a detailed comparison among these three formalisms, see Ref. \cite{das2023finite}. Numerous studies have employed TFD to investigate various physical processes \cite{ruiz2024exploring, cabral2023compton, ferreira2022tfd, correa2023aether, cabral2024lorentz}, including one within the GEM framework \cite{evangelista2025gravitational}, where the thermal contribution appears to enhance the interaction. This highlights the importance of including thermal effects in physical models, particularly for understanding processes in extreme environments such as stellar cores. Here, the main objective is to present a detailed analysis of Lorentz-violating gravitational \( e^- + e^+ \to \ell^- + \ell^+ \) scattering, considering both zero and finite temperature scenarios.

The paper is organized as follows. In Sec. \ref{sec2}, we present the Gravitoelectromagnetism (GEM) formalism and highlight its key properties. Section \ref{sec3} focuses on the effects of Lorentz violation on the scattering process: Subsec. \ref{sec31} introduces the Lorentz-violating (LV) term, and Subsec. \ref{sec32} applies the formalism to the gravitational \( e^- + e^+ \to \ell^- + \ell^+ \)   scattering. In Sec. \ref{sec4}, we extend the analysis to finite temperature. Subsection \ref{sec41} provides an overview of TFD, and Subsec. \ref{sec42} incorporates thermal effects into the Lorentz-violating gravitational scattering. Finally, in Sec. \ref{con}, we present our concluding remarks.

\section{Gravitoelectromagnetism: Theoretical Definitions}\label{sec2}

In this section, we discuss the theoretical foundations of Gravitoelectromagnetism (GEM), a theory that aims to describe gravity in a manner analogous to electromagnetism. GEM is valid only in the weak-field limit, making it unsuitable for describing strong-field objects such as black holes and pulsars \cite{bakopoulos2016gravitoelectromagnetism}. To formulate GEM, three common approaches are typically considered: (i) based on the similarity between the linearized Einstein field equations in the weak-field limit and Maxwell's equations \cite{mashhoon2003gravitoelectromagnetism}, {known as the Linearized GEM}, (ii) decomposing the Weyl tensor into its gravitoelectric and gravitomagnetic components \cite{faqir2010lagrangian}, {also referred to as the Weyl GEM}, and (iii) conducting an analysis based on the tidal tensor \cite{costa2008gravitoelectromagnetic}.

In this work, we focus on the second approach, as it provides a well-established Lagrangian formulation that facilitates quantization, as discussed in the literature \cite{faqir2010lagrangian}. In this framework, the Weyl tensor, defined as
{\small
\begin{eqnarray}
    C_{\alpha\sigma\mu\nu} &=& R_{\alpha\sigma\mu\nu} - \frac{1}{2} \left( R_{\nu\alpha} g_{\mu\sigma} + R_{\mu\sigma} g_{\nu\alpha} - R_{\nu\sigma} g_{\mu\alpha} - R_{\mu\alpha} g_{\nu\sigma} \right)+ \frac{1}{6} R \left( g_{\nu\alpha} g_{\mu\sigma} - g_{\nu\sigma} g_{\mu\alpha} \right),
\end{eqnarray}}
where \( R_{\alpha\sigma\mu\nu} \) is the Riemann tensor, \( R_{\nu\alpha} \) is the Ricci tensor, \( R \) represents the Ricci scalar, and \( g_{\mu\nu} \) is the metric tensor, is decomposed into its gravitoelectric and gravitomagnetic components as follows:  
\begin{eqnarray}
 {E}_{ij} &=& -C_{0i0j}, \\
   {B}_{ij} &=& \frac{1}{2} \epsilon_{ikl} C^{kl}_{\phantom{kl}0j}, 
\end{eqnarray}
with \( i, j, k, \dots = 1,2,3 \). { Indeed, these fields are generally defined in terms of the Weyl tensor as
${E}_{\mu\nu} = u^\kappa u^\lambda C_{\mu\kappa\nu\lambda}$ and
${B}_{\mu\nu} = \tfrac{1}{2} u^\kappa u^\lambda \epsilon_{\alpha\rho\kappa\mu} C^{\alpha\rho}\,_{\nu\lambda}$,
where $u^\lambda$ denotes the four-velocity of an observer \cite{Campbell, Ber}. By adopting a local rest frame defined by $u^\lambda$, one finds that only the spatial components of ${E}_{\mu\nu}$ and ${B}_{\mu\nu}$ remain nonvanishing. Moreover, these tensors are trace-free, so that each possesses five independent components. Together, they account for the ten independent components of the Weyl tensor. This decomposition allows one to analyze gravitational tidal effects and spacetime curvature in a manner analogous to the electric and magnetic fields in electromagnetism.} From this, the GEM, or Maxwell-like, equations for flat spacetime with a source term can be expressed as:

\begin{eqnarray}
    \partial^{i} {E}^{ij} &=& 4\pi G \rho^j, \\
    \partial^{i} {B}^{ij} &=& 0, \\
    \epsilon^{\langle ikl \rangle} \partial^k {B}^{lj} - \frac{1}{c} \frac{\partial{E}^{ij}}{\partial t} &=& \frac{4\pi G}{c} J^{ij}, \\
    \epsilon^{\langle ikl \rangle} \partial^k {E}^{lj} + \frac{1}{c} \frac{\partial {B}^{ij}}{\partial t} &=& 0,
\end{eqnarray}
where \( \rho^j \) is a mass vector density, \( J^{ij} \) represents a mass current tensorial density, \( \epsilon^{\langle \dots \rangle} \) is the Levi-Civita symbol, \( c \) is the speed of light, and \( G \) is the gravitational constant. {In the Weyl GEM formalism, the definition of the Maxwell-like equations focuses on the properties of the dynamical fields and waves of the interaction, analogous to the $E$ and $B$ fields in electromagnetism.} Furthermore, the fields \( {E}^{ij} \) and \( {B}^{ij} \) are symmetric and defined as
\begin{eqnarray}
    {E} &=& -\text{grad}\;\varphi - \frac{1}{c} \frac{\partial \tilde{A}}{\partial t}, \\  
    {B} &=& \text{curl}\;\tilde{A},  
\end{eqnarray}
where \( \varphi \) represents the potential vector, analogous to the scalar potential $\phi$ in electromagnetism. Furthermore, \( \tilde{A} \) represents the potential tensor, with components \( A^{ij} \). Note that the quantities introduced here are analogous to those in electromagnetism, except that they all have an additional rank. This higher-order structure is necessary to describe a spin-2 particle. With these definitions, we can now introduce the GEM field tensor, given by 
\begin{eqnarray}
    \mathcal{F}^{\mu\nu\alpha}=\partial^\mu A^{\nu\alpha}-\partial^\nu A^{\mu\alpha},\label{tensorfield}
\end{eqnarray}
and the GEM dual tensor, written as  
\begin{eqnarray}
    \mathcal{G}^{\mu\nu\alpha}=\frac{1}{2}\epsilon^{\mu\nu\gamma\sigma} \eta^{\alpha\xi} \mathcal{F}_{\gamma\sigma\xi}.
\end{eqnarray}
Thus, the GEM field equations can be rewritten in covariant form as
\begin{eqnarray}
     \partial_{\mu} \mathcal{F}^{\mu\nu\alpha}&=&-\frac{4\pi G}{c}\mathcal{J}^{\nu\alpha},\label{tensorF}\\
    \partial_\mu \mathcal{G}^{\mu\langle\nu\alpha\rangle}&=&0 ,\label{tensorG}
\end{eqnarray}
where $\mathcal{J}^{\nu\alpha}$ is a second-order tensor that depends on the mass density vector $\rho^j$ and the mass current density $J^{ij}$. From these equations, the GEM Lagrangian is defined as
\begin{eqnarray}
   \mathcal{L}_{\text{GEM}}=-\frac{1}{16\pi}\mathcal{F}_{\mu\nu\alpha}\mathcal{F}^{\mu\nu\alpha}-G\mathcal{J}^{\nu\alpha} A_{\nu\alpha}.
\end{eqnarray}
When considering the interaction between gravitons and fermions, the total Lagrangian can be expressed as
\begin{eqnarray}
    \mathcal{L}_{\text{Total}}=&-&\frac{1}{16\pi }\mathcal{F}_{\mu\nu\alpha}\mathcal{F}^{\mu\nu\alpha}-\frac{i}{2}(\Bar{\psi}\gamma^\mu \partial_\mu\psi-\partial_\mu\Bar{\psi}\gamma^\mu\psi)\nonumber\\
    &+&m\Bar{\psi}\psi-\frac{i\kappa}{4}A_{\mu\nu}(\Bar{\psi}\gamma^\mu\partial^\nu\psi-\partial^\mu\Bar{\psi}\gamma^\nu\psi),\label{LTOTAL}
\end{eqnarray}
where \( \psi \) is the Dirac field, with \( \Bar{\psi} = \psi^\dagger \gamma_0 \), \( \gamma^\mu \) represent the Dirac matrices, \( m \) is the fermion mass, and \( \kappa = \sqrt{8\pi G} \) is the coupling constant.

Now, let us compare this approach, where the potential tensor is the primary quantity, with the linearized Einstein equation expressed in terms of the metric perturbation 
$h_{\mu\nu}$. In contrast to the latter, which involves a perturbative expansion of the metric around flat spacetime, the first approach {uses $A_{\mu\nu}$ as a fundamental field}, offering a different perspective on how gravity can be described. Coupling Eq. (\ref{tensorfield}) with Eq. (\ref{tensorF}) yields
\begin{eqnarray}
    \Box A^{\nu\alpha}-\partial^\nu(\partial_\mu A^{\mu\alpha})=4\pi G\mathcal{J}^{\nu\alpha}.
\end{eqnarray}
Due to the gauge transformation of the GEM potential, which is analogous to the electromagnetic case, we can apply the Lorentz-like gauge condition \( \partial_\mu A^{\mu\alpha}=0 \) and rewrite the previous equation as  
\begin{eqnarray}
    \Box A^{\nu\alpha}=4\pi G \mathcal{J}^{\nu\alpha}.
\end{eqnarray}
This condition closely resembles the linearized Einstein field equation expressed in terms of a perturbation of the metric, given by
\begin{eqnarray}
    \Box \Bar{h}_{\mu\nu}=16\pi G T_{\mu\nu},
\end{eqnarray}
with \( \Bar{h}_{\mu\nu}=h_{\mu\nu}-\frac{1}{2}\eta_{\mu\nu}h \) \cite{misner1973k}. This similarity reinforces the connection between Gravitoelectromagnetism and General Relativity. Moreover, from this comparison, we recognize the significance of the tensor \( A^{\nu\alpha} \) in {the Weyl GEM formalism}, as it encapsulates all the essential information required for describing the theory. Notably, the potential tensor arises naturally in {this formulation of GEM} and is directly related to the description of the gravitational field, rather than being associated with a perturbation of the metric in the weak-field limit, {as is the case in the linearized GEM representation}.
 Consequently, the Lagrangian formulation presented in this study facilitates the quantization process, providing an interpretation of the graviton as a spin-2 particle. For a more detailed discussion on the differences between GEM and General Relativity, see Refs. \cite{bakopoulos2016gravitoelectromagnetism, mashhoon2003gravitoelectromagnetism, alesandrogravitacional}. In the next section, we will use the introduced quantities to examine the effects of Lorentz violation on gravitational \( e^- + e^+ \to \ell^- + \ell^+ \)  scattering. Specifically, we will analyze how the interaction between fermions, mediated by gravitons, is modified in the presence of a background Lorentz-violating field.

\section{Effects of Lorentz violation on gravitational $e^-+e^+\to \ell^-+\ell^+$ scattering at zero temperature}\label{sec3}

In this section, we analyze gravitationall \( e^- + e^+ \to \ell^- + \ell^+ \) scattering in the presence of Lorentz violation (LV) in the fermion-antifermion interaction Lagrangian. Specifically, we examine how a Lorentz-violating background field modifies the process when mediated by gravitons.

To establish a solid foundation for this discussion, we first introduce the Lorentz-violating term incorporated into the gravitational sector. Originally formulated within the CPT-odd sector of the Standard-Model Extension (SME) framework \cite{bailey2015short}, this dimension-5 background field modifies the dynamics of interacting particles, potentially leading to deviations from standard gravitational interactions. Its structure resembles other non-minimal CPT-odd terms in the QED sector \cite{kostelecky2019gauge, borges2022external}. Given that GEM provides an electromagnetic analogy, this choice is particularly convenient. By examining the effects and properties of this LV coefficient, we aim to quantify its impact on the scattering amplitude and cross-section, which will be discussed in the following subsections.

\subsection{The Lorentz-Violating Background Field}\label{sec31}

\begin{figure}[!h]
	\centering
	\includegraphics[width=0.8\linewidth]{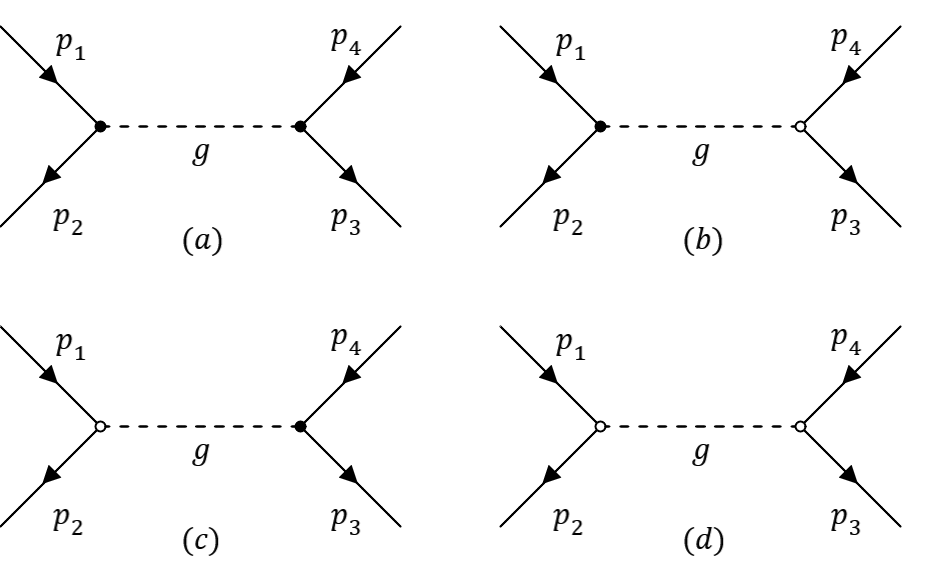}
	\caption{Feynman diagram for $e^-+e^+\to \ell^- +\ell^+$ scattering in the presence of Lorentz violation. The Lorentz-preserving diagram is shown in (a), while diagrams (b) and (c) feature a single Lorentz-violating vertex. The diagram in (d) includes Lorentz violation at both vertices.}
	\label{feynmandiag}
\end{figure}

The gravitational sector of the Lorentz-violating Standard-Model Extension (SME) has been extensively studied in the literature \cite{kostelecky2004gravity, bailey2015short, kostelecky2016testing} and incorporates a wide range of terms. In this work, we focus on the fifth-order Lorentz-violating coefficient  $\mathcal{K}^{(5)}_{\mu\nu\alpha\iota\phi}$, which has {dimension of inverse of mass}. Lorentz violation is introduced through a nonminimal coupling in the vertex interaction between fermions and gravitons \cite{alesandrogravitacional}. Since the free Lagrangian remains unaltered, the graviton propagator is not modified. Then, the Lagrangian given in (\ref{LTOTAL}) can be rewritten as
\begin{eqnarray}
        \mathcal{L}_{\text{Total}} = -\frac{1}{16\pi} \mathcal{F}_{\mu\nu\alpha} \mathcal{F}^{\mu\nu\alpha} - \bar{\psi} (i\gamma^\mu \overset{\leftrightarrow}{D_\mu} - m ) \psi,
\end{eqnarray}
where the covariant derivative, modified by the insertion of the LV field, is defined as
\begin{eqnarray}
    \overset{\leftrightarrow}{D_\mu} = \overset{\leftrightarrow}{\partial_\mu} - \frac{1}{2} \kappa A_{\mu\nu} \overset{\leftrightarrow}{\partial^\nu} + \frac{i}{4} \mathcal{K}^{(5)}_{\mu\nu\alpha\iota\phi} \gamma^\nu \mathcal{F}^{\alpha\iota\phi}.
\end{eqnarray}
Using the covariant derivative defined above, the total Lagrangian takes the form
{\small
\begin{eqnarray}
    \mathcal{L}_{\text{Total}}&=&-\frac{1}{16\pi} \mathcal{F}_{\mu\nu\alpha} \mathcal{F}^{\mu\nu\alpha} - \frac{i}{2}(\Bar{\psi}\gamma^\mu\partial_\mu\psi - \partial_\mu\Bar{\psi}\gamma^\mu\psi) + m\Bar{\psi}\psi \nonumber \\
    &-& \frac{i\kappa}{4} A_{\mu\nu} (\Bar{\psi}\gamma^\mu\partial^\nu\psi - \partial^\mu\Bar{\psi}\gamma^\nu\psi)+\frac{i}{4}\mathcal{K}^{(5)}_{\mu\nu\alpha\iota\phi}\mathcal{F}^{\alpha\iota\phi}\bar{\psi}\sigma^{\mu\nu}\psi,\nonumber\\
\end{eqnarray}}
where $\sigma^{\mu\nu}=\frac{i}{2}(\gamma^\mu\gamma^\nu-\gamma^\nu\gamma^\mu)$. The vertices associated with the Feynman diagram for gravitational $e^-+e^+\to \ell^-+\ell^+$ scattering can now be expressed as  
\begin{eqnarray}
	\bullet &\to& V_0^{\mu\nu}=-\frac{i\kappa}{4}(\gamma^\mu p_1^\nu+p_2^\mu\gamma^\nu);\\
	\circ &\to& V_1^{\mu\nu}=\frac{i}{2}\mathcal{K}_{(5)}^{\iota\phi\alpha\mu\nu}\sigma_{\iota\phi}q_\alpha;\label{Vertex}
\end{eqnarray}
where $\bullet$ represents the Lorentz-preserving vertex and $\circ$ denotes the Lorentz-violating one. Furthermore, $p_1\;(p_2)$ denotes the momentum of the fermion (antifermion), while $q_\alpha=(\sqrt{s},0,0,0)$ represents the momentum contribution from particle conservation. The Feynman diagram incorporating all interaction contributions is shown in Figure \ref{feynmandiag}.

{Let us analyze the Lorentz-violating (LV) Lagrangian density. The LV contribution of the Lagrangian is
	\begin{align}
		\mathcal{L}_{LV}=\frac{i}{4}\mathcal{K}^{(5)}_{\mu\nu\alpha\iota\phi}\mathcal{F}^{\alpha\iota\phi}\bar{\psi}\sigma^{\mu\nu}\psi,
	\end{align}
	where we see that it has mass dimension $d=5$ ($\bar{\psi}\sigma^{\mu\nu}\psi=3$, $\mathcal{F}^{\alpha\iota\phi}=2$). This implies that the LV coefficient has dimensions of inverse mass. Such a structure is very similar to the Lagrangian $\mathcal{L}^{(5)}_{\psi F}$ of Table I in Ref. \cite{kostelecky2019gauge}, where a LV coefficient $H_F^{(5)\mu\nu\alpha\beta}$ is considered in QED. This suggests that adopting an analogous structure here is appropriate if we aim to describe gravity in a way analogous to electromagnetism. Furthermore, the GEM tensor is antisymmetric in the first two indices, as are the Pauli matrices ($\mathcal{F}^{\alpha\iota\phi}=-\mathcal{F}^{\iota\alpha\phi}$ and $\sigma^{\mu\nu}=-\sigma^{\nu\mu}$). From this, we can extract information about the LV coefficient:
	\begin{align}
		&\mathcal{K}^{(5)}_{\mu\nu\alpha\iota\phi}=-\mathcal{K}^{(5)}_{\mu\nu\iota\alpha\phi}\quad\text{(antisymmetric in $\alpha,\iota$ due to $\mathcal{F}^{\alpha\iota\phi}$),}\\
		&\mathcal{K}^{(5)}_{\mu\nu\alpha\iota\phi}=-\mathcal{K}^{(5)}_{\nu\mu\alpha\iota\phi}\quad\text{(antisymmetric in $\mu,\nu$ due to the Dirac matrices $\sigma^{\mu\nu}$).}
	\end{align}}
	To facilitate working with the Lorentz-violating vertex, we can redefine this coefficient in terms of the Minkowski metric, the Levi-Civita symbol, and a vector, {analogously to the proposal in \cite{borges2022external}}, i.e.,
	\begin{eqnarray}
		{\mathcal{K}^{(5)}_{\iota\phi\alpha\mu\nu}=\eta_{\alpha[\iota}\epsilon_{\phi]\lambda\mu\nu}K^{(5)\lambda},}
	\end{eqnarray}
{where the brackets denote the antisymmetric indices of $\mathcal{K}^{(5)}$. Here, we rename the LV coefficient indices to facilitate writing the interaction vertex. For simplicity, we denote $K^{(5)\lambda}=K^{\lambda}$. The expression above also preserves antisymmetry in $\alpha$ and $\mu$, allowing us to express the LV coefficient as
		\begin{align}
			\mathcal{K}^{(5)}_{\iota\phi\alpha\mu\nu}=\frac 12 (\eta_{\alpha [ \iota}\epsilon_{\phi ]\lambda\mu\nu}-\eta_{\mu [ \iota}\epsilon_{\phi ]\lambda\alpha\nu})K^\lambda. \label{Antissym}
	\end{align}}
	Based on these considerations, we can now redefine the Lorentz-violating vertex in (\ref{Vertex}), where the violation explicitly depends on the {vector $K^\lambda$}. Expanding the antisymmetrization in \eqref{Antissym}, we obtain the following LV vertex:
	\begin{eqnarray}
		{V_1^{\mu\nu}=\frac i8 (\eta_{\alpha  \rho}\epsilon_{\sigma \lambda\mu\nu}-\eta_{\alpha\sigma}\epsilon_{\rho\lambda\mu\nu}-\eta_{\mu\rho}\epsilon_{\sigma\lambda\alpha\nu}+\eta_{\mu\sigma}\epsilon_{\rho\lambda\alpha\nu})\mathrm{K}^\lambda q^\alpha \sigma^{\rho\sigma}.}
	\end{eqnarray}

In the next subsection, we will provide a comprehensive discussion of the Lorentz-violating gravitational $e^-+e^+\to\ell^-+\ell^+$ scattering and compute the corrections to the cross section up to second order in the Lorentz-violating terms.

\subsection{Lorentz-violating cross section at zero temperature}\label{sec32}

Here, we investigate the cross section for gravitational $e^-+e^+\to \ell^- +\ell^+$ scattering  in the presence of Lorentz violation. The differential cross section is defined as
\begin{eqnarray}
     \frac{d\sigma}{d\Omega}=\frac{1}{64\pi^2s}\left\langle\left|\mathcal{M}\right|^2\right\rangle,\label{cross}
\end{eqnarray}
with $\mathcal{M}$ being the transition amplitude. To proceed with the calculations, we choose the center-of-mass (CM) frame, where the four-momenta are given by
\begin{eqnarray}
    p_1&=&(E,0,0,-E),\\
    p_2&=&(E,0,0,E),\\
    p_3&=&(E,E\sin{\theta},0,E\cos{\theta}),\\
    p_4&=&(E,-E\sin{\theta},0,-E\cos{\theta}).
\end{eqnarray}
From these four-momenta, the Mandelstam variables are given by
\begin{eqnarray}
    s=4E^2;\quad t=-2E^2(1+\cos\theta);\quad u=-2E^2(1-\cos\theta).\nonumber\\ \label{Mandelstam}
\end{eqnarray}
The interaction Lagrangian describing this process is given by
{\small
\begin{eqnarray}
    \mathcal{L}_{\text{I}}=- \frac{i\kappa}{4} A_{\mu\nu} (\Bar{\psi}\gamma^\mu\partial^\nu\psi - \partial^\mu\Bar{\psi}\gamma^\nu\psi)+\frac{i}{4}\mathcal{K}_{(5)}^{\iota\phi\alpha\mu\nu}\mathcal{F}_{\alpha\mu\nu}\bar{\psi}\sigma_{\iota\phi}\psi.\nonumber\\
\end{eqnarray}}
The transition amplitude, on the other hand, is defined as 
\begin{eqnarray}
    \mathcal{M}=\bra{f}\hat{S}^{(2)}\ket{i},\label{M}
\end{eqnarray}
where $S^{(2)}$ is the second-order scattering matrix, expressed as  
\begin{eqnarray}
    \hat{S}^{(2)}=-\frac{1}{2}\int\int d^4x\;d^4y\;\tau[\mathcal{L}_{\text{I}}(x)\mathcal{L}_{\text{I}}(y)],
\end{eqnarray}
with $\tau$ denoting the time-ordering operator. The asymptotic states $\ket{f}$ and $\ket{i}$ are given as  
\begin{eqnarray}
    \ket{i}&=&b^\dagger_{p_1}c^\dagger_{p_2}\ket{0},\\
    \ket{f}&=&b^\dagger_{p_3^\prime}c^\dagger_{p_4^\prime}\ket{0}.
\end{eqnarray}
Since Lorentz violation is introduced in the interaction term, the graviton propagator remains unaltered. Thus, it is given by  
\begin{eqnarray}
    \Delta_{\mu\nu\xi\rho}^{(0)}(q)=\frac{i}{2q^2}(\eta_{\mu\xi}\eta_{\nu\rho}+\eta_{\mu\rho}\eta_{\nu\xi}-\eta_{\mu\nu}\eta_{\xi\rho}).
\end{eqnarray}
The total scattering amplitude is obtained by summing over all the contributions shown in Figure (\ref{feynmandiag}). The expected result consists of the standard (non-LV) gravitational $e^-+e^+\to\ell^-+\ell^+$ scattering plus a correction due to Lorentz violation. The last diagram, (d), is not considered, as Lorentz-violating effects are small and this contribution is of higher order in Lorentz-violation corrections.

With this in mind, we now combine all previous definitions to compute the transition amplitude for the graviton-fermion interaction. This quantity is expected to be
\begin{eqnarray}
    \mathcal{M}=\mathcal{M}_{0}+\mathcal{M}_{1},  
\end{eqnarray}
where $\mathcal{M}_{0}$ represents the non-LV contribution, while $\mathcal{M}_{1}$ corresponds to the LV correction. Explicitly, the total transition amplitude can be written as
\begin{eqnarray}
    \mathcal{M}&=&-\frac{i}{2q^2}\big( \eta_{\mu\xi} \eta_{\nu\rho} + \eta_{\mu\rho} \eta_{\nu\xi} - \eta_{\mu\nu} \eta_{\xi\rho} \big)\nonumber\\
    &\times&\big[\bar{v}(p_1) V_0^{\mu\nu} u(p_2) \bar{u}(p_3) V_0^{\xi\rho} v(p_4)\nonumber\\
    &+&\bar{v}(p_1) V_{0}^{\mu\nu} u(p_2) \bar{u}(p_3) V_{1}^{\xi\rho} v(p_4)\nonumber\\
    &+&\bar{v}(p_1) V_{1}^{\mu\nu} u(p_2) \bar{u}(p_3) V_0^{\xi\rho} v(p_4)\big],
\end{eqnarray}
where the first term corresponds to diagram (a), while the second and third terms represent diagrams (b) and (c), respectively, as shown in Figure \ref{feynmandiag}.

We are now in a position to compute the probability amplitude, given by
\begin{eqnarray}
     \left\langle \left| \mathcal{M} \right|^2 \right\rangle=\frac{1}{4}\sum_{\text{spins}}|\mathcal{M}|^2= \left\langle \left| \left(\mathcal{M}_{0}+\mathcal{M}_{1}\right) \right|^2 \right\rangle.\label{Probamplitude}
\end{eqnarray}
Using the completeness relation, the summation over spin states can be expressed as
\begin{eqnarray} \sum_{\text{spin}} u(p,s) \bar{u}(p,s) &=& \slashed{p}_i + m \quad (fermions), \\ \sum_{\text{spin}} v(p,s) \bar{v}(p,s) &=& \slashed{p}_i - m \quad (antifermions), \end{eqnarray}
where $\slashed{p}_i=\gamma_\mu p_i^\mu$. Applying the trace properties, we obtain
{\small
\begin{eqnarray} 
&& \big(\bar{v}(p_1) V_0^{\mu\nu} u(p_2) \bar{u}(p_3) V_0^{\xi\rho} v(p_4)\big) \big(\bar{v}(p_4) V_0^{\delta\theta} u(p_3) \bar{u}(p_2) V_0^{\lambda\zeta} v(p_1)\big) \nonumber\\ &&= \Tr\big\{(\slashed{p}_1 - m_e) V_0^{\mu\nu} (\slashed{p}_2 + m_e) V_0^{\lambda\zeta} \big\}\nonumber\\
&&\times\Tr\big\{(\slashed{p}_3 + m_\mu) V_0^{\xi\rho} (\slashed{p}_4 - m_\mu) V_0^{\delta\theta} \big\}. 
\end{eqnarray}}
This result allows us to compute the squared transition amplitude in the ultrarelativistic limit. Here, we adopt the coordinate choice $K_\lambda=(K_t,0,0,K_z)$ for the LV coefficient, which allows the transition amplitude to be written as
\begin{eqnarray} 
	\left\langle \left| \mathcal{M} \right|^2 \right\rangle=\frac{1}{128}\kappa^2 U^4[\kappa^2(\cos 2\theta+3)+K_z^2(7\cos 2\theta+9)].\label{Mprob} 
\end{eqnarray}
It is important to note that the appearance of the $K_z^2$ term in the squared transition amplitude constitutes the Lorentz-violating contribution, which arises from the chosen background configuration $K_\lambda = (K_t, 0, 0, K_z)$ and introduces a preferred direction along the $z$-axis. Furthermore, the Lorentz-violating effect reaches its maximum at $\theta = 0$ and its minimum at $\theta = \pi/2$. It is important to note that if the Lorentz-violation coefficient is set to zero, i.e., $K_z \to 0$, we recover the standard gravitational  $e^-+e^+\to\ell^-+\ell^+$ scattering process in the absence of Lorentz violation. In this case, the transition amplitude simplifies to

\begin{eqnarray}
    \left\langle \left| \mathcal{M} \right|^2 \right\rangle= \left\langle \left| \mathcal{M}_0 \right|^2 \right\rangle = \frac{1}{128}\kappa^4 U^4(\cos 2\theta+3).
\end{eqnarray}
Using this result, we substitute Eq. (\ref{Mprob}) into Eq. (\ref{cross}) and compute the differential cross section, which yields
\begin{eqnarray}
        {\bigg(\frac{d\sigma}{d\Omega}\bigg)=\frac{1}{64\pi^2 s}\bigg\{\frac{1}{128}\kappa^2 U^4[\kappa^2(\cos 2\theta+3)+K_z^2(7\cos 2\theta+9)]\bigg\},}\nonumber\\
\end{eqnarray}
which can be integrated to obtain the total cross section, given by
\begin{eqnarray}
     {\sigma=\frac{\kappa^4E^2}{3072\pi}\bigg(1+\frac{5}{2}\frac{K_z^2}{\kappa^2}\bigg).}
\end{eqnarray}
Furthermore, by setting $K_z\to0$, we obtain the gravitational cross section for $e^-+e^+\to\ell^-+\ell^+$ scattering, given by
\begin{eqnarray}
     \sigma_{\text{GEM}}=\frac{\kappa^4 E^2}{3072\pi}.\label{crosszero}
\end{eqnarray}
Thus, we can express the Lorentz-violating cross section as
\begin{equation}
   {\sigma=\sigma_{\text{GEM}}	\bigg(1+\frac{5}{2}\frac{K_z^2}{\kappa^2}\bigg).}\label{crosseven}
\end{equation}
As a result, we explicitly observe the contribution of the Lorentz-violating term in comparison to the non-LV cross section. These modifications affect the energy dependence of the gravitational cross section, potentially leading to observable signatures if gravitons are ever detected. However, a direct comparison with experimental data is not possible, as no such data are currently available.

To establish a meaningful comparison between GEM and QED, it is essential to recognize a fundamental difference in their coupling constants. In GEM, the coupling constant \(\kappa\) has the dimension of inverse energy, approximately given by \(\kappa \approx 1/M_p\), where \(M_p\) is the Planck mass. In contrast, the fine-structure constant in QED,  \(\alpha = e^2/4\pi\epsilon_0\), is a dimensionless quantity. To bridge this difference, we introduce a characteristic energy scale  \(E_c\) for the scattering process and redefine the coupling constant as $ \kappa \to \kappa^\prime = \kappa E_c.$
This rescaling renders \(\kappa^\prime\) dimensionless, allowing for a more direct analogy between GEM and QED interactions.

In the next section, we explore Lorentz-violating scattering at finite temperature using the TFD formalism.

\section{Lorentz-violating effects in gravitational $e^-+e^+\to \ell^-+\ell^+$ scattering at finite temperature}\label{sec4}

In this section, we investigate Lorentz-violating $e^- + e^+ \to \ell^- + \ell^+$ scattering at finite temperature. To incorporate thermal effects, we adopt the Thermo Field Dynamics (TFD) formalism, which provides a mathematically convenient framework, particularly at tree level.

In the following subsections, we first introduce the fundamental concepts of the TFD formalism and then apply them to the scattering process under study.

\subsection{The TFD formalism}\label{sec41}

TFD is a quantum field theory framework that incorporates thermal effects in both equilibrium and non-equilibrium systems \cite{Umezawa2, arimitsu1987non}.  Applying TFD requires two fundamental elements. The first is the doubling of the Hilbert space into non-tilde and tilde components, expressed as $\mathcal{H}_T = \mathcal{H} \otimes \tilde{\mathcal{H}}$. Here, $\mathcal{H}$, known as the non-tilde space, represents the usual Hilbert space, while $\tilde{\mathcal{H}}$, the tilde space, is an auxiliary (doubled) Hilbert space. The second essential element is the Bogoliubov transformation.

In the TFD formalism, an arbitrary operator, say $Q$, follows the tilde conjugation rules:
\begin{eqnarray}
    &&(Q_iQ_j)^\til=\Tilde{Q}_i\Tilde{Q}_j;\\
    &&(cQ_i+Q_j)^\til=c^*\Tilde{Q}_i+\Tilde{Q}_j;\\
    &&(Q_i^\dagger)^\til=(\Tilde{Q}_i)^\dagger;\\
    &&(\Tilde{Q}_i)^\til=\pm Q_i;
\end{eqnarray}
where $c$ is a complex number, and  the sign $(+)$ is used for bosons, while $(-)$ is used for fermions.

Due to the doubling of the Hilbert space, we can define a thermal vacuum state $\ket{0(\beta)}$ as a superposition of the usual vacuum states $\ket{0}$ from quantum mechanics, where $\beta=1/k_B T$. This thermal state is given by  
\begin{eqnarray}
    \ket{0(\beta)}=\frac{1}{\sqrt{Z(\beta)}}\sum_n e^{-\beta E_n/2}\ket{n,\tilde{n}},
\end{eqnarray}
where $Z(\beta)$ is the partition function. Additionally, based on the Bogoliubov transformations, the field operators are redefined through a rotation in the doubled space. These transformations can be expressed as
\begin{eqnarray}
    b_{s,p} &=& U(\beta)b_{s,p}(\beta)+V(\beta)\Tilde{b}_{s,p}^\dagger(\beta),\nonumber\\
    \Tilde{b}_{s,p} &=& U(\beta)\Tilde{b}_{s,p}(\beta)-V(\beta)b_{s,p}^\dagger(\beta),\nonumber\\
    b_{s,p}^\dagger &=& U(\beta)b_{s,p}^\dagger(\beta)+V(\beta)\Tilde{b}_{s,p}(\beta),\nonumber\\
    \Tilde{b}_{s,p}^\dagger &=& U(\beta)\Tilde{b}_{s,p}^\dagger(\beta)-V(\beta)b_{s,p}(\beta),\label{boguferm}
\end{eqnarray}
for fermions, where $U(\beta)=\cos\theta(\beta)$ and $V(\beta)=\sin\theta(\beta)$. Similarly, for bosons we have
\begin{eqnarray}
    a_{\lambda,k} &=& U^\prime(\beta)a_{\lambda,k}(\beta)+V^\prime(\beta)\Tilde{a}_{\lambda,k}^\dagger(\beta),\nonumber\\
    \Tilde{a}_{\lambda,k} &=& U^\prime(\beta)\Tilde{a}_{\lambda,k}(\beta)+V^\prime(\beta)a_{\lambda,k}^\dagger(\beta),\nonumber\\
    a_{\lambda,k}^\dagger &=& U^\prime(\beta)a_{\lambda,k}^\dagger(\beta)+V^\prime(\beta)\Tilde{a}_{\lambda,k}(\beta),\nonumber\\
    \Tilde{a}_{\lambda,k}^\dagger &=& U^\prime(\beta)\Tilde{a}_{\lambda,k}^\dagger(\beta)+V^\prime(\beta)a_{\lambda,k}(\beta),\label{bogubos}
\end{eqnarray}
where $U^\prime(\beta)=\cosh\theta(\beta)$ and $V^\prime(\beta)=\sinh\theta(\beta)$. Here, $p(k)$ and $s(\lambda)$ represent the momenta and the spins (polarizations) of fermions (bosons), respectively. Note that the operators presented in Eq. (\ref{boguferm}) and (\ref{bogubos}) are the same as those found in the fermionic and GEM fields, respectively.

In the next subsection, the tools presented here will be used to investigate the thermal corrections to electron-positron scattering in the presence of Lorentz-violating terms.

\subsection{Lorentz-violating cross section at finite temperature}\label{sec42}

Here, the Lorentz-violating gravitational $e^-+e^+\to \ell^-+\ell^+$ scattering at finite temperature will be addressed. The definitions introduced in the previous subsection, derived from TFD, will be applied to express the transition amplitude as
\begin{eqnarray}
    \mathcal{M}(\beta) = \bra{f(\beta)}\hat{S}^{(2)}\ket{i(\beta)},\label{MB}
\end{eqnarray}
where the asymptotic states are now given as
\begin{eqnarray}
    \ket{i(\beta)} &=& b^\dagger_{p_1}c^\dagger_{p_2}\ket{0(\beta)},\\
    \ket{f(\beta)} &=& b^\dagger_{p_3^\prime}c^\dagger_{p_4^\prime}\ket{0(\beta)}.
\end{eqnarray}
Due to the doubling of the Hilbert space, the interaction Lagrangian is given by the sum of the non-tilde and tilde components, namely,
\begin{eqnarray}
    \hat{\mathcal{L}}_{\text{I}}(x) &=& \mathcal{L}_{\text{I}}(x) - \Tilde{\mathcal{L}}_{\text{I}}(x) \\
    &=& -\frac{i\kappa}{4}A_{\mu\nu}(\bar{\psi}\gamma^\mu\partial^\nu\psi\nonumber\\ 
    &-& \partial^\mu\bar{\psi}\gamma^\nu\psi) + \frac{i\kappa}{4}\tilde{A}_{\mu\nu}(\tilde{\bar{\psi}}\gamma^\mu\partial^\nu\tilde{\psi} - \partial^\mu\tilde{\bar{\psi}}\gamma^\nu\tilde{\psi})\nonumber\\
    &+&\frac{i}{4}\mathcal{K}_{(5)}^{\iota\phi\alpha\mu\nu}\mathcal{F}_{\alpha\mu\nu}\bar{\psi}\sigma_{\iota\phi}\psi-\frac{i}{4}\mathcal{K}_{(5)}^{\iota\phi\alpha\mu\nu}\tilde{\mathcal{F}}_{\alpha\mu\nu}\tilde{\bar{\psi}}\sigma_{\iota\phi}\tilde{\psi}.\nonumber
\end{eqnarray}

The graviton propagator at finite temperature\footnote{For details on the calculation of the graviton propagator at finite temperature, see Ref. \cite{graviton}.} is given by 
\begin{eqnarray}
    \Delta_{\mu\nu\xi\rho}(q,\beta) &=& \Delta_{\mu\nu\xi\rho}^{(0)}(q) + \Delta_{\mu\nu\xi\rho}^{(\beta)}(q) \nonumber\\
    &=& \left(\eta_{\mu\xi}\eta_{\nu\rho} + \eta_{\mu\rho}\eta_{\nu\xi} - \eta_{\mu\nu}\eta_{\xi\rho}\right)\nonumber\\
   &\times& \Bigg[\frac{i}{2q^2} - \frac{2\pi i \delta(q^2)}{e^{\beta q_0} - 1}
    \begin{pmatrix}
        1 & e^{\frac{\beta q_0}{2}} \\
        e^{\frac{\beta q_0}{2}} & 1
    \end{pmatrix}\Bigg].
\end{eqnarray}

The cross section will have a structure similar to that of the zero-temperature case, where
\begin{eqnarray}
     \frac{d\sigma(\beta)}{d\Omega}=\frac{1}{64\pi^2s}\left\langle\left|\mathcal{M}(\beta)\right|^2\right\rangle.\label{crossbeta}
\end{eqnarray}

With this approach, we can directly proceed with the calculation of the probability amplitude at finite temperature, leading to the following result
{\small
\begin{eqnarray}
   && \left\langle \left| \mathcal{M}(\beta) \right|^2 \right\rangle=\bigg[{\frac{1}{128}\kappa^2 U^4[\kappa^2(\cos 2\theta+3)+K_z^2(7\cos 2\theta+9)]}\bigg]\nonumber\\
     &\times&\Bigg[\frac{(\tanh(\beta E/2)+1)}{2}\Bigg]^4\Bigg(\frac{256 \pi^2 \delta^2(s) E^4}{(e^{2\beta E}-1)^2}+1\Bigg).
\end{eqnarray}}
Then, the differential cross section given in Eq. (\ref{crossbeta}) becomes
{\small
\begin{eqnarray}
   && \bigg(\frac{d\sigma(\beta)}{d\Omega}\bigg)=\frac{1}{64\pi^2 s}\bigg[{\frac{1}{128}\kappa^2 U^4[\kappa^2(\cos 2\theta+3)+K_z^2(7\cos 2\theta+9)]}\bigg]\nonumber\\
     &\times&\Bigg[\frac{(\tanh(\beta E/2)+1)}{2}\Bigg]^4\Bigg(\frac{256 \pi^2 \delta^2(s) E^4}{(e^{2\beta E}-1)^2}+1\Bigg).
\end{eqnarray}}
Using the Dirac delta property $x\delta(x)=0$, we obtain
{\small
\begin{eqnarray}
   \bigg(\frac{d\sigma(\beta)}{d\Omega}\bigg)=\frac{1}{64\pi^2 s}\mathcal{D}(\beta)\bigg[{\frac{1}{128}\kappa^2 U^4[\kappa^2(\cos 2\theta+3)+K_z^2(7\cos 2\theta+9)]}\bigg],
\end{eqnarray}}
where, for simplicity, has been defined  
\begin{eqnarray}
    \mathcal{D}(\beta)=\Bigg[\frac{(\tanh(\beta E/2)+1)}{2}\Bigg]^4.
\end{eqnarray}
From this, we can calculate the total cross section at finite temperature, given by
\begin{eqnarray}
    \sigma(\beta)=\sigma_{\text{GEM}}(\beta)\bigg({1+\frac{5}{2}\frac{K_z^2}{\kappa^2}}\bigg)\mathcal{D}(\beta).
\end{eqnarray}

Note that in the zero-temperature limit, i.e., $\beta\to \infty$, or, $T\to0$, we obtain $\mathcal{D}(\beta)\to 1$, recovering the Lorentz-violating scattering at zero temperature, as given in Eq. (\ref{crosseven}). Furthermore, if we take $\beta \to \infty$ and $K_z\to 0$, we retrieve the standard gravitational scattering at zero temperature without the Lorentz-violating contribution, as presented in Eq. (\ref{crosszero}). Therefore, extending this analysis to include thermal effects and Lorentz violation provides a more comprehensive understanding of how this interaction manifests in nature, where temperature plays a fundamental role. Additionally, within the framework of the SME theory, these extensions offer deeper insights into how symmetry-breaking mechanisms influence the dynamics of the process, potentially revealing new physical effects that would be absent in a purely zero-temperature or Lorentz-invariant scenario.

\vspace{0.2cm}

\section{Conclusion}\label{con}

The Standard Model, despite its remarkable success, does not account for gravity, highlighting the need for theoretical extensions such as the SME. In this work, we investigated the gravitational sector of the SME using the GEM framework, which models gravity in analogy with electromagnetism. The theory's Lagrangian formulation, based on the potential tensor $A_{\mu\nu}$ , supports applications to various physical processes, including scattering phenomena. We examined the gravitational $e^-+e^+\to\ell^-+\ell^+$  scattering process, focusing on how fermions interact through graviton exchange. Lorentz violation was introduced via a fifth-order background field from the nonminimal SME gravity sector, structurally similar to CPT-odd terms in the QED sector.  We derived a modified cross section that includes a Lorentz-violating additive contribution.

To incorporate thermal effects, we employed the TFD formalism. This approach introduces temperature dependence by doubling the Hilbert space and redefining the field operators through Bogoliubov transformations. As a result, the gravitational interaction acquires explicit thermal corrections, and the cross section becomes a function of the temperature. An asymptotic analysis of this thermal contribution reveals that increasing temperature tends to enhance the scattering amplitude.

These results are particularly relevant from a fundamental physics perspective. Lorentz violation and finite temperature are both expected to play significant roles in regimes approaching the Planck scale, where quantum gravity effects become important. Therefore, exploring their combined influence on physical processes contributes to our broader understanding of how spacetime symmetries and thermodynamics may interplay in the search for a unified theory at very high energies, such as those found in the early universe or extreme astrophysical environments.

\section*{Acknowledgments}

This work by A. F. S. is partially supported by National Council for Scientific and Technological Develo\-pment - CNPq project No. 312406/2023-1. L. A. S. E. thanks CAPES for financial support. The authors thank Professor Pablo Rodrigo Alves de Souza for his valuable initial discussions.

\section*{Data Availability Statement}

No Data associated in the manuscript.


\global\long\def\link#1#2{\href{http://eudml.org/#1}{#2}}
 \global\long\def\doi#1#2{\href{http://dx.doi.org/#1}{#2}}
 \global\long\def\arXiv#1#2{\href{http://arxiv.org/abs/#1}{arXiv:#1 [#2]}}
 \global\long\def\arXivOld#1{\href{http://arxiv.org/abs/#1}{arXiv:#1}}

\end{document}